\documentclass[12pt,russian]{article}
 \usepackage{helvet}
 \usepackage[T1]{fontenc}
 \usepackage[cp1251]{inputenc}
 \usepackage{geometry}
 \usepackage{graphicx}
 \usepackage{setspace}
 \usepackage{babel}

 \renewcommand{\vec}[1]{\mbox{\boldmath $#1$}}

 \def\gsim{\lower.4ex\hbox{$\;\buildrel >\over{\scriptstyle\sim}\;$}}
 \def\lsim{\lower.4ex\hbox{$\;\buildrel <\over{\scriptstyle\sim}\;$}}
 
 \def\bl{\par\vskip 12pt\noindent}
 \def\bll{\par\vskip 24pt\noindent}

\begin{document}

\vskip 1.5 cm

{\large Zh. Exp. Teor. Fiz. {\bf 169}, iss. 3 (2026)}

\bll

\begin{center}
{\large{\bf Small-Scale Dynamo for Full Spectrum of Hydrodynamic\\[0.3 truecm] Turbulence in Kazantsev Model}}
\end{center}

\bl

\centerline{L.\,L.~Kitchatinov\footnote{E-mail: kit@iszf.irk.ru}}

\bl

\begin{center}
{\it Institute of Solar-Terrestrial Physics, Lermontov Str. 126A, Irkutsk, 664033, Russia}
\\
\bl
Received 2 Febuary 2026 / Accepted 7 February 2026
\end{center}

\bl
\hspace{0.8 truecm}
\parbox{14.4 truecm}{
{\bf Abstract.} A method is proposed for computing coefficients in the Kazantsev equation of small-scale dynamo for the full spectrum of hydromagnetic turbulence comprising the inertial range together with the range of viscous dissipation. The dynamo equation with so-defined coefficients is solved numerically for magnetic (Rm) and hydrodynamic (Re) Reynolds numbers from $10^2$ to $10^8$. The threshold value ${\rm Rm}_c$ for onset of dynamo increases initially with Re but then saturates at a constant value of ${\rm Rm}_c \simeq 300$ for ${\rm Re}\gsim 10^5$. In the case of small Prandtl number Pm = Rm/Re << 1, the field growth rate is also small and depends logarithmically on Rm. In this case, the magnetic energy spectrum peaks around the scale of Ohmic dissipation, which decreases with increasing Pm. The decrease stops at the scale of viscous dissipation while the growth rate increases sharply when Pm approaches the value of one. The increase in the growth rate proceeds to ${\rm Pm} > 1$ but slows down and then saturates at a value somewhat below the inverse lifetime of most short-living eddies. An explanation of the results is proposed.

\bll


 }
 
\bll

\reversemarginpar

\setlength{\baselineskip}{0.7 truecm}

\noindent{\bf 1. Introduction}

\noindent
Magnetic field generation by flows of electrically conducting fluids - the hydromagnetic dynamo - is a conventional explanation for the origin of magnetic fields in astrophysics [\ref{ZRS83}-\ref{RKH13}]. The presence of turbulence driven by various instabilities catalyses dynamos. Large and small-scale dynamos can be distinguished by comparing with the (external) scale of turbulence. The large-scale planetary, stellar or galactic dynamos are controlled by rotation and rotationally induced violation of mirror-reflection parity in the tur\-bu\-lence [\ref{ZRS83},\ref{KR84}]. Small-scale dynamos do not require rotation. 

The physics of small-scale dynamos may seem simple: field-line stretching by turbu\-lence under conservation of mass inside the magnetic flux-tubes amplifies the magnetic energy. The simplicity is however illusive. The field amplification is accompanied by spectral transport of magnetic energy to small scales where Ohmic dissipation eventually comes into play. Rates of field amplification and scale decrease equal in their order of magnitude. The problem of small-scale dynamo therefore demands a quanti\-ta\-tive comparison of these two competing processes. The small-scale dynamo equation derived by Kazantsev [\ref{K67}] includes the field amplification parameter and the scale-dependent diffusion standing for the two above-mentioned processes. Solution of the Kazantsev equation for a turbulent spectrum model [\ref{NRS83}] confirmed the possibility of small-scale dynamos. 

Later on, the Kazantsev model has been applied to study various aspects of hydro\-mag\-ne\-tic dynamos [\ref{Cea99}-\ref{MYS25}]. The model was also generalised to account for the finite correlation time of the turbulence and for the turbulence deviation from Gaussian statistics [\ref{Kea22},\ref{Kea24}]. Nevertheless, the model's simplicity and clarity of the physics on which it is based retain the Kazantsev model as the principal tool of the small-scale dynamo theory. 

Uncertainty regarding the dependence of the field amplification rate [\ref{MB10},\ref{KR12}] or characteristic scales of the generated fields [\ref{Nea22}] on key model parameters remain among unsolved problems of the theory. This is possibly caused by difficulties in the specification of coefficients in the Kazantsev equation. The coefficients include the correlation function of  turbulent velocities. The correlation function cannot be prescribed arbitrarily. It should correspond to a positive-definite energy spectrum [\ref{MY67}] (the rather popular \lq lon\-gi\-tu\-di\-nal' correlation function $T_{LL}(r) = v\ell (1 - r^2/\ell^2)/3$ corresponds to a sign-altering spectrum).  
An integral transform of a spectrum into a correlation function is known [\ref{MY67}], but its application to the energy spectrum $E(k) \propto k^{-5/3}$ is problematic. We shall see also that the main contribution to the field amplification coefficient in the Kazantsev equation comes from the viscous dissipation range of the energy spectrum, which range, therefore, should be accounted for. In these circumstances, the small-scale dynamo problems are  treated either by using a model spectrum, which allows an analytical transformation into a correlation function, or by using asymptotic methods to prove a presence of growing solutions in the inertial scale range. 

This paper proposes a numerical method for conversion of the kinetic energy spec\-trum, including the dissipative range, into the correlation function. The method allows computing the two coefficients in the Kazantsev equation for Reynolds numbers ${\rm Re} \leq 10^8$. Solving the eigenvalue problem with so-defined coefficients gives the growth rates and eigenfunctions for the kinematic small-scale dynamo problem in dependence on the usual and magnetic Reynolds numbers. Numerical transform of the eigenfunctions gives in turn the magnetic energy spectra. Note that a computation of the scale-dependent diffusion in the Kazantsev equation has its own interest for it defines the separation rate of close particles in a turbulent flow also [\ref{VK86},\ref{K89}].
 \bl
 \noindent{\bf 2. Model}

\noindent
The Kazantsev model is formulated for the case of a magnetic field in isotropic statis\-ti\-cally steady and homogeneous turbulence with delta-correlated in time velocity field (the last assumptions may be excessive [\ref{VK86}]).
In this case, the correlation tensor of the magnetic field ${\vec b}({\vec r},t)$,
\begin{equation}
    \langle b_i({\vec r}_1,t)b_j({\vec r}_2,t)\rangle =
    B_{LL}\frac{r_ir_j}{r^2} +
    \frac{1}{2r}\frac{\partial(r^2B_{LL})}{\partial r}
    \left(\delta_{ij} - \frac{r_ir_j}{r^2}\right),
    \label{1}
\end{equation}
is uniquely defined by the so-called longitudinal correlation function 
$B_{LL}(r,t)$ [\ref{MY67}], which depends on the distance $r =
\mid{\vec r}_1 - {\vec r}_2\mid$ and time $t$ only. The dynamo equation is written for this function:
\begin{equation}
    \frac{\partial B_{LL}}{\partial t} =
    \frac{2}{r^4}\frac{\partial}{\partial r}r^4
    \left(\eta_{_{\rm T}}(r) + \eta\right)\frac{\partial B_{LL}}{\partial r}
    + Q(r)B_{LL} .
    \label{2}
\end{equation}
In this equation, $\eta$ is the magnetic diffusivity, $\eta_{_{\rm T}}(r)$ is the scale-dependent turbulent diffusion which, together with the field amplification coefficient $Q(r)$, is defined in terms of the correlation function $T_{LL}$ of the turbulent flow:
\begin{eqnarray}
    \eta_{_{\rm T}}(r) &=& T_{LL}(0) - T_{LL}(r)
    \nonumber \\
    Q(r) &=& -\frac{2}{r^4}\frac{{\rm d}}{{\rm d}r}r^4
    \frac{{\rm d}T_{LL}}{{\rm d} r}.
    \label{3}
\end{eqnarray}
More spcifically, $T_{LL}(r)$ is the time-integrated correlation function of turbulent velocity $\vec u$:
\begin{eqnarray}
    T_{LL}(r) &=& \int\limits_0^\infty U_{LL}(r,\tau){\rm d}\tau ,
    \label{4}
    \\
    U_{LL}(r,\tau) &=& r_ir_j\langle u_i({\vec r}_1,t)
    u_j({\vec r}_1 + {\vec r},t+\tau)\rangle /r^2.
    \nonumber
\end{eqnarray}

The scale-dependent diffusion (\ref{3}) has a clear meaning. The correlation function $T_{LL}(r)$ is contributed by fluctuations of scales $\gsim r$. The difference $T_{LL}(0) - T_{LL}(r)$, therefore, gives the diffusion due to fluctuations of scales $\lsim r$. The coefficient $Q(r)$ accounts for the magnetic field amplification due to field-line stretching by fluctuations of these latter scales. 

If the functions (\ref{3}) were known, a numerical solution of the one-dimensional in space equation (\ref{2}) would not present any difficulties. 
The problem is to define these functions only. The function $T_{LL}(r)$
is related to the corresponding spectrum $W(k)$ by the integral transform
[\ref{MY67}]
\begin{equation}
    T_{LL}(r) = \int\limits_0^\infty W(k)g(kr)\frac{{\rm d}k}{(kr)^2},
    \label{5}
\end{equation}
where
\begin{equation}
    g(kr) = \frac{\sin (kr)}{kr} - \cos (kr)
    \label{6}
\end{equation}
and $W(k) = E(k)\tau_k$ is the product of the turbulent velocity spectrum
$E(k)$
\begin{equation}
    \langle u^2\rangle = \int\limits_0^\infty E(k){\rm d}k
    \label{7}
\end{equation}
and the life-time $\tau_k$ of turbulent eddies of scale $1/k$.

Computation with Eq. (\ref{5}) followed by a derivation of the coefficients (\ref{3}) is not an optimal procedure, however. Computation of the generation coefficient $Q(r)$ is simplified by the fact that $g(kr)/(kr)^2$ in Eq. (\ref{6}) is an eigenfunction of the diffusion operator:
\begin{equation}
    \frac{1}{r^4}\frac{\partial}{\partial r}r^4\frac{\partial}{\partial r}
    \frac{g(kr)}{(kr)^2} = -k^2\frac{g(kr)}{(kr)^2}.
    \label{8}
\end{equation}
The coefficient $Q(r)$ will therefore be computed with the equation
\begin{equation}
    Q(r) = \frac{2}{r^2}\int\limits_0^\infty W(k)g(kr)\,{\rm d}k .
    \label{9}
\end{equation}
The uncertainty for small $r$ can be fixed with the equation
\begin{equation}
    g(z) = z^2(1 - z^2(1-z^2/28)/10)/3 + O(z^8) .
    \label{10}
\end{equation}
It, in particular, follows from this equation that
\begin{equation}
    Q(0) = \frac{2}{3}\int\limits_0^\infty W(k)k^2\,{\rm d}k.
    \label{11}
\end{equation}

The quantity $Q(0)$ is present in the equation for magnetic energy density, 
\begin{equation}
    \frac{\partial\langle b^2\rangle}{\partial t} =
    30\eta B^{\prime\prime}_{LL}(0) + Q(0)\langle b^2\rangle ,
    \label{12}
\end{equation}
where $B^{\prime\prime}_{LL}(0)$ means the second spatial derivative at $r = 0$. It follows from this equation that the positive-definite quantity $Q(0)$ is the rate of energy transmission to the magnetic field. The first term on the right side of Eq. (\ref{12}) accounts for the Ohmic dissipation. 

An important conclusion follows from the fact that the integral in Eq.\,(\ref{11}) diverges for the power spectrum $E(k) \propto
k^{-5/3}$ at large $k$. This means that a computation of $Q(r)$ of Eq.\,(\ref{9}) has to allow for the viscous dissipation range in the turbulence spectrum and the largest rate of magnetic energy feed settles in this range. It does not mean, however, that the magnetic energy spectrum does also have a maximum in the viscous dissipation range. The Ohmic dissipation can prevent this. We shall see that the maximum in the magnetic spectrum shifts in the direction of decreasing wave numbers with decreasing ${\rm Pm} = \nu /\eta$. 

Searching for a solution of the dynamo equation (\ref{2}) in the form of $B_{LL}(r,t) = \exp(\gamma t) B(r)/3$ leads to the eigenvalue problem, 
\begin{equation}
    \gamma B(r) =  \frac{2}{r^4}\frac{\partial}{\partial r}r^4
    \left(\eta_{_{\rm T}}(r) + \eta\right)\frac{\partial B(r)}{\partial r}
    + Q(r)B(r),
    \label{13}
\end{equation}
for computing the growth rate $\gamma$ of magnetic energy and correlation function $B(r)$ of the magnetic field. Only the dominant eigenmodes corresponding to the largest growth rate are considered in what follows. Boundary conditions for Eq.\,(\ref{13}) are the regularity of the correlation function at $r=0$, $B^\prime(0) = 0$, and a fall of the function to zero for large $r$. 

The turbulence spectrum in the inertial range is known [\ref{K41}]:
$E(k) = 2C\epsilon^{2/3}k^{-5/3}$ ($\epsilon$ is the energy injection rate per unit mass and $C \simeq 3/2$). The characteristic velocity of turbulent eddies of scale $1/k$ is $v_k = (\epsilon/k)^{1/3}$ and the eddy's lifetime $\tau_k = 1/(kv_k) = (\epsilon k^2)^{-1/3}$. Within the inertial range, we therefore have 
\begin{equation}
    W(k) = 3\,\epsilon^{1/3}k^{-7/3} .
    \label{14}
\end{equation}

\begin{figure}[htb]
 \begin{center}
 \includegraphics[width=12.0 cm]{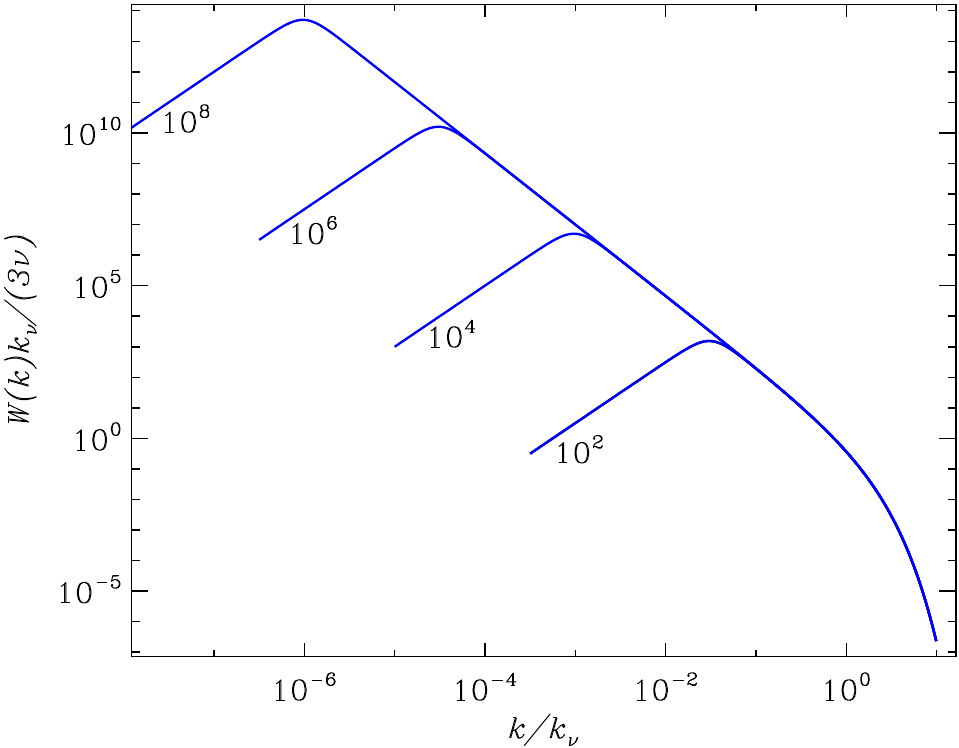}
 \end{center}
 \begin{description}
 \item{\small {\bf Fig.\,1.} Spectra $W(k)$ of Eq.\,(\ref{16}) for several Reynolds number values, which values mark their corresponding lines.
    }
 \end{description}
\end{figure}

The inertial range of wave numbers has an upper bound $k_\nu = (\epsilon/\nu^3)^{1/4}$ where the eddy's turnover time $\tau_k$ equals the viscous dissipation time $1/(k^2\nu )$. The range of $k \sim k_\nu$ is important for small-scale dynamos. We therefore rewrite Eq.\,(\ref{14}) as
\begin{equation}
    W(k) = 3\frac{\nu}{k_\nu}(k/{k_\nu})^{-7/3},
    \label{15}
\end{equation}
where the spectrum's dimension is absorbed by the coefficient $\nu /k_\nu$.

In the dissipation range, $k \gsim k_\nu$, the spectrum changes to exponential dependence on the wave number [\ref{Mea97}]. The spectrum (\ref{15}) can be extended into this range by multiplying with $\exp(-k/k_\nu)$. The low bound $k_0 = 1/\ell_0$ of the inertial range is the inverse of the spatial scale $\ell_0$ of energy injection. The spectrum has a maximum around $k \approx k_0$ and decreases with decreasing $k$ for smaller wave numbers. The Reynolds number ${\rm Re} = \ell_0 v_{k_0}/\nu$ can be expressed in terms of the scale ratio,
${\rm Re} = (k_\nu/k_0)^{4/3}$. This paper applies the \lq full spectrum' model 
\begin{equation}
    W(k) = 3\frac{\nu}{k_\nu}\,\frac{(k/k_\nu)^n\exp(-k/k_\nu)}
    {{\rm Re}^{-3p/4} + (k/k_\nu)^p},\ \ p = 7/3 + n.
    \label{16}
\end{equation}
The model reproduces the spectrum (\ref{15}) in the inertial range, $k_0 < k < k_\nu$, the exponential law in the dissipation range, and $W(k) \propto k^n$ for $k < k_0$ (n=2 in computa\-ti\-ons to follow). The spectra for several values of the Reynolds number are shown in Fig.\,1. 
 \bl
 \noindent{\bf 3. Numerical method}

\noindent
Finite-difference grids in either radius $r$ and wave number $k$ were used. In both variables, the grid points cover a range from zero to a maximum value equal $r_{\rm max} = 100\ell_0$ for radius and $k_{\rm max} = 10 k_\nu$ for wave number. Grids cannot be uniform in so broad ranges. 

\begin{figure}[htb]
 \begin{center}
 \includegraphics[width=12.0 cm]{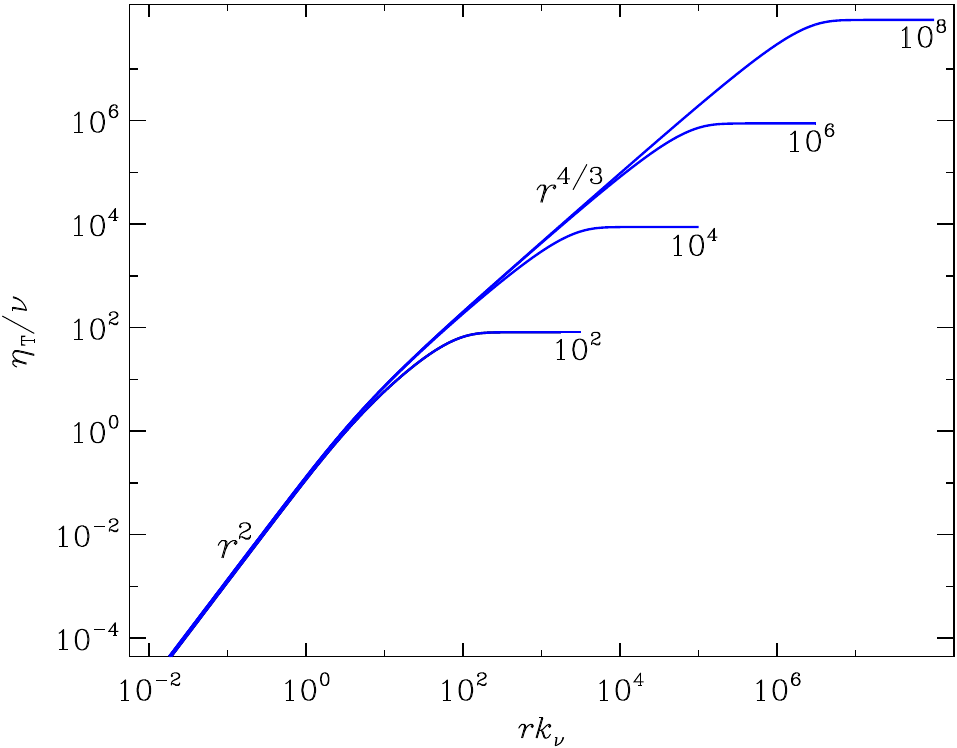}
 \end{center}
 \begin{description}
 \item{\small {\bf Fig.\,2.} The scale-dependent diffusion coefficients $\eta_{_{\rm T}}(r)$. Numbers near the lines show the corresponding Reynolds number.
    }
 \end{description}
\end{figure}

The second grid point in the wave number $k_2 = \delta k_0$ ($k_1 = 0$), where
$\delta \ll 1$ is the model parameter. The further grid points are given by iterations,  
\begin{equation}
    k_{m+1} - k_m = (k_m - k_{m-1})(1+\varepsilon ),
    \label{17}
\end{equation}
where $\varepsilon \ll 1$ is a small parameter as well. The relative distance between grid points,
\begin{equation}
    (k_{m+1} - k_m)/k_m = \delta k_0/k_m + \varepsilon ,
    \label{18}
\end{equation}
approaches $\varepsilon$ with increasing $m$. The iterations stop upon exceeding the maximum wave number $k_{\rm max}$.

Similarly, for the grid in $r$, $r_1 = 0$, $r_2 = \delta/k_\nu$ and further on according to the iteration equation
\begin{equation}
    r_{m+1} - r_m = (r_m - r_{m-1})(1 + \varepsilon )
    \label{19}
\end{equation}
until the value of $r_{\rm max}$ is exceeded. The computations were done with $\delta =\varepsilon = 0.01$. This permits computations with a moderate number of grid points (<2000) even for the largest ${\rm Re} = 10^8$ of our computations. 

\begin{figure}[htb]
 \begin{center}
 \includegraphics[width=12.0 cm]{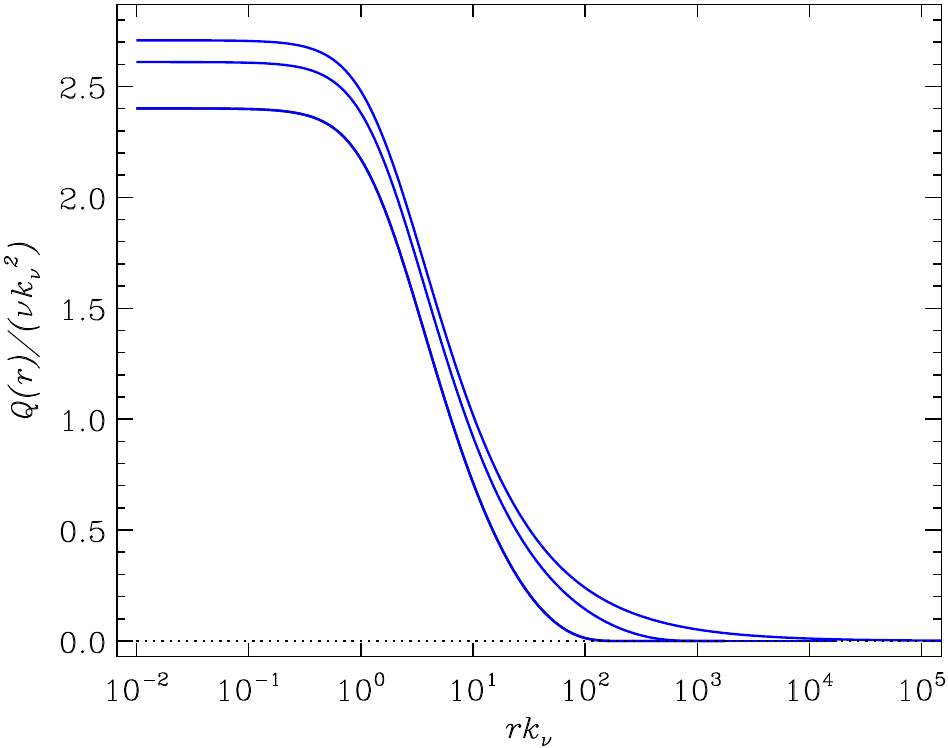}
 \end{center}
 \begin{description}
 \item{\small {\bf Fig.\,3.} Normalised field amplification coefficient $Q$ (\ref{9}) in dependence on the relative scale $rk_\nu$.
     The three profiles correspond to different Reynolds numbers that vary as $10^2$, $10^3$ and $10^8$ from the lower to the upper line.
    }
 \end{description}
\end{figure}

One more comment on the computation method is in order. In spite of the relative small distance between the grid points, the function (\ref{6}) can vary strongly between neighbouring grid points if its argument is large. Integration in Eq.\,(\ref{9}) and computation of a similar integral for the diffusion coefficient is permitted by the fact that the oscillating trigonometric functions enter the integrals in multiplication with smooth functions that vary little between the neighbouring grid points. These smooth functions were interpolated with cubic spline. Inter-grid integrals of the products of the trigonometric functions with cubic spline are derived analytically and expressed in terms of the spline coefficients. The upper limit in integrals of Eqs (\ref{5}) and (\ref{9}) was put equal to $k_{\rm max}$. 

The scale-dependent diffusivity computed with this method is shown in Fig.\,2. Three scale ranges can be distinguished. In the viscous dissipation range, $rk_\nu < 1$, inhomogeneity of turbulent flow is close to linear dependence on $r$. Therefore, $\eta_{_{\rm T}} \propto r^2$ in this range (it can be shown that $\eta(r) \simeq r^2Q(0)/20$ in this range). In the inertial range of $1 < rk_\nu < {\rm Re}^{3/4}$, computations reproduce the Richardson law, $\eta_{_{\rm T}} \propto r^{4/3}$ [\ref{R26},\ref{O41}], for separation of close particles in turbulent flow. For $r > \ell_0$, the diffusivity eventually approaches the constant value of $\eta_{_{\rm T}} \simeq \ell_0v_{k_0}/3$ for large-scale fields. 

The amplification coefficient $Q(r)$ is shown in Fig.\,3. As the Reynolds number increases, the normalised coefficient $Q(r)/\nu k_\nu^2$ approaches rapidly a universal profile whose value is small for $rk_\nu \gg 1$.

\begin{figure}[htb]
 \begin{center}
 \includegraphics[width=12.0 cm]{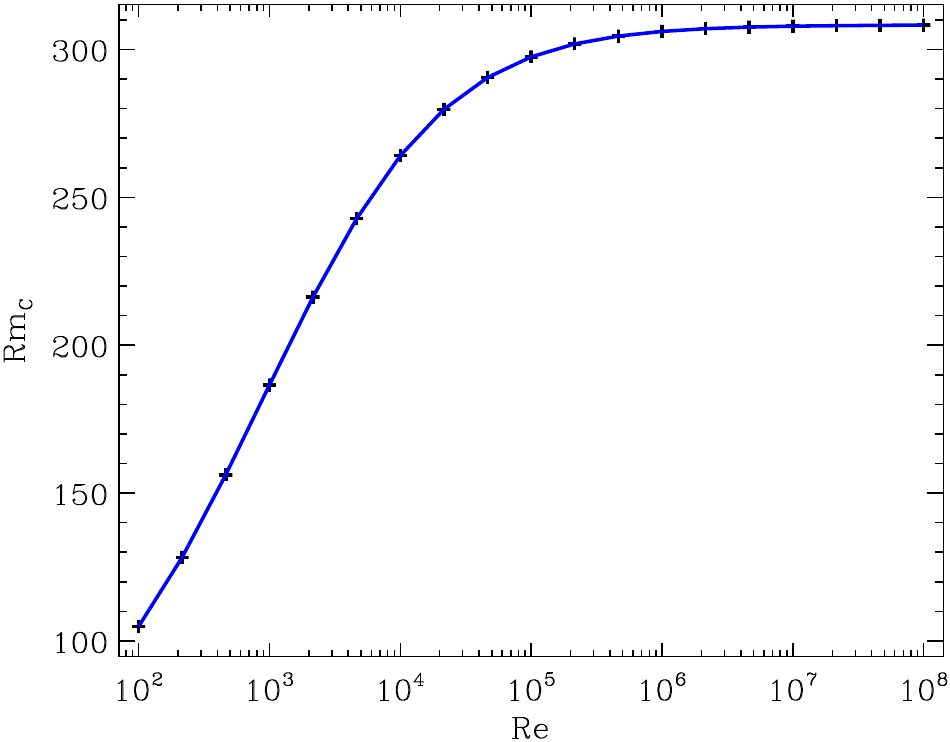}
 \end{center}
 \begin{description}
 \item{\small {\bf Fig.\,4.} Threshold value ${\rm Rm_c}$ for onset of the small-scale dynamo as the function of Reynolds number. The computed points are shown by crosses.
    }
 \end{description}
\end{figure}

The eigenvalue equation (\ref{13}) was solved by the inverse iteration method with second-order accurate finite difference approximation of derivatives. The end-point boundary condition was applied at $r_{\rm max}$: $B(r_{\rm max}) = 0$. The eigenfunction of the linear equation was normalised by the condition 
$B(0) = 1$.
 \bl
 \noindent{\bf 4. Results and discussion}

\noindent
Growing solutions with $\gamma > 0$ emerge for the magnetic Reynolds number ${\rm Rm} = \ell_0v_{k_0}/\eta$ exceeding a certain threshold value $\rm Rm_c$. 
The dependence of $\rm Rm_c$ on hydrodynamical Reynolds number is shown in Fig.\,4. 

\begin{figure}[htb]
 \begin{center}
 \includegraphics[width=12.0 cm]{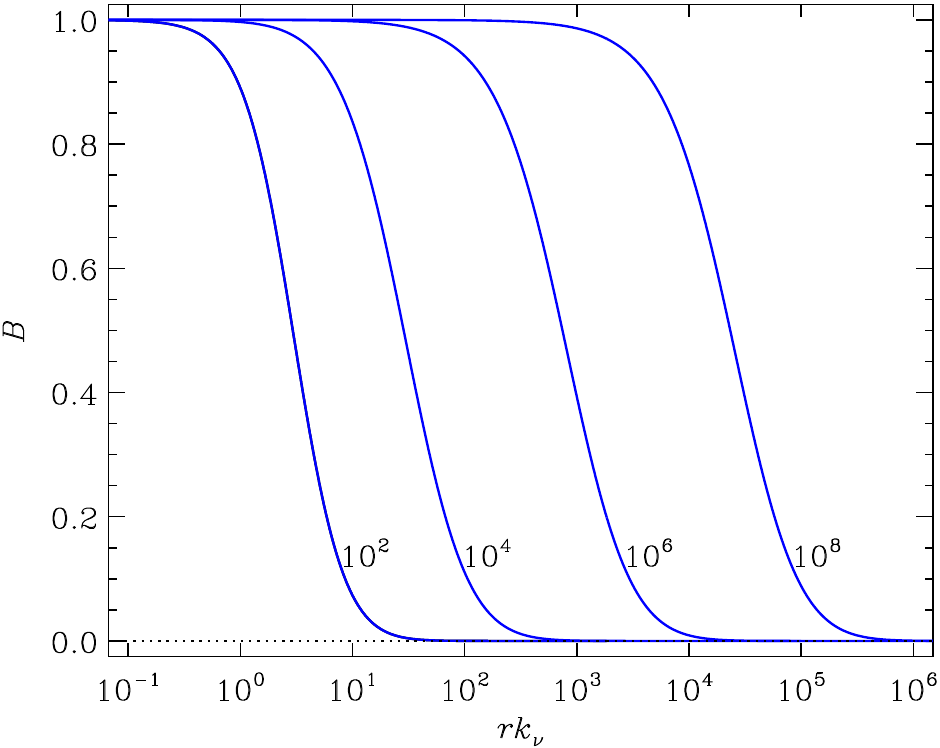}
 \end{center}
 \begin{description}
 \item{\small {\bf Fig.\,5.} Eigenfunctions of the dynamo equation (\ref{13}) for the threshold magnetic Reynolds number $\rm Rm_c$. The lines are marked with the corresponding Reynolds number values. 
    }
 \end{description}
\end{figure}

Different statements on the character of this dependence can be met in the literature. A steady increase of $\rm Rm_c$ with ${\rm Re}$ was predicted in [\ref{BC04}]. It can be seen in Fig.\,4 that the initial increase saturates at the level of ${\rm Rm_c} \simeq 300$ but the saturation onsets at large ${\rm Re} \geq 10^6$ not accessible for 3D numerical experiments. A possibility of  saturation in the increase of $\rm Rm_c$ was discussed in [\ref{Sea04},\ref{Iea07}].

\begin{figure}[htb]
 \begin{center}
 \includegraphics[width=12.0 cm]{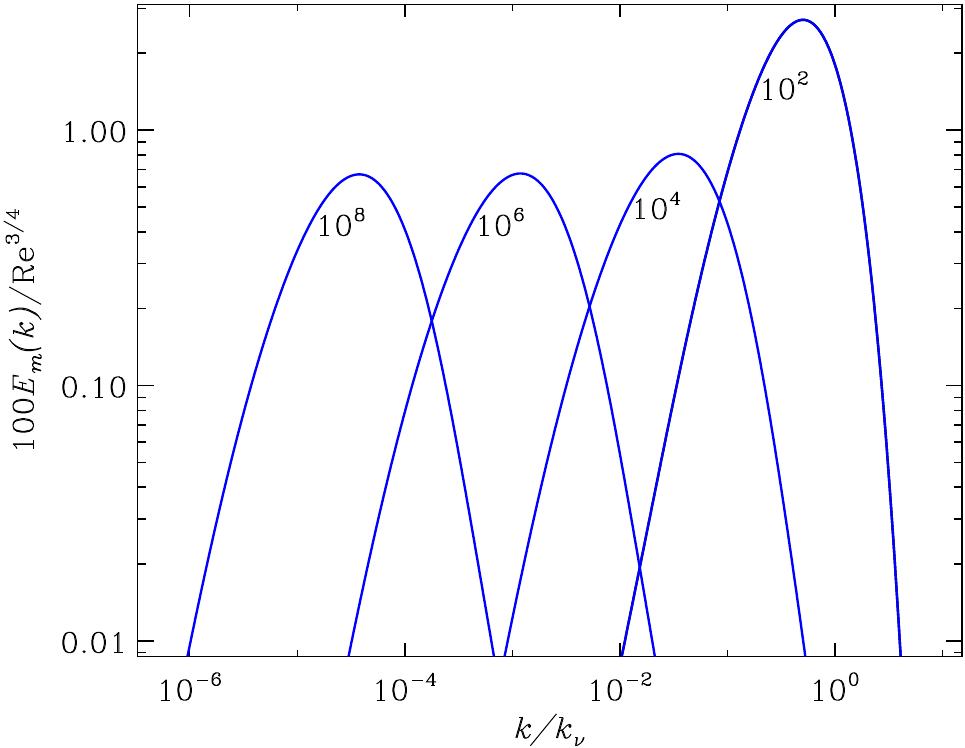}
 \end{center}
 \begin{description}
 \item{\small {\bf Fig.\,6.} Magnetic energy spectra of Eq.\,(\ref{20}) for marginal dynamos. Lines are marked by their corresponding Reynolds number values. The coefficient $100/{\rm Re}^{3/4}$ is introduced for the convenience of showing spectra for different $\rm Re$ on the same plot. 
    }
 \end{description}
\end{figure}

Correlation functions $B(r)$ for marginal dynamos are shown in Fig.\,5. It can be seen from this Figure that magnetic field correlation length increases relative to the viscous dissipation scale with increasing $\rm Re$. This can be explained by the decrease in ${\rm Pm} = {\rm Rm_c}/{\rm Re}$ and the resulting Ohmic dissipation enhancement. In parallel, the spectral region of magnetic field generation shifts to larger scales. This is evident from the shape of magnetic energy spectra, 
\begin{equation}
    E_m(k) = \frac{2}{3\pi}\int\limits_0^{r_{\rm max}}
    B(r)g(kr)(kr)^2\,{\rm d}r ,
    \label{20}
\end{equation}
shown in Fig.\,6 (numerical integration in Eq.\,(\ref{20}) is done by the method explained above). In the saturation region of Fig.\,4, the position of maximum in the spectra obeys the relation $k/k_\nu \propto {\rm Pm}^{3/4}$. This means that the field amplification for ${\rm Pm} < 1$ localises at the Ohmic dissipation scale $k_\eta \simeq (\epsilon /\eta^3)^{1/4}$ in the inertial range.

\begin{figure}[htb]
 \begin{center}
 \includegraphics[width=12.0 cm]{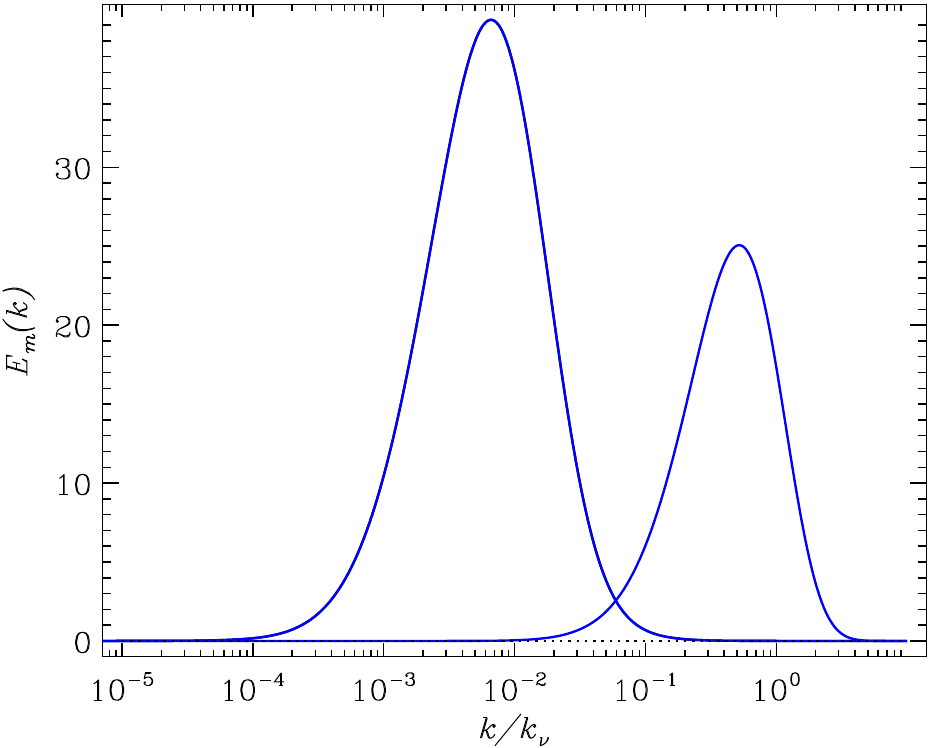}
 \end{center}
 \begin{description}
 \item{\small {\bf Fig.\,7.} Magnetic energy spectra for ${\rm Re} = 10^5$. {\sl Left}: marginally stable mode for the threshold magnetic Reynolds number ${\rm Rm_c} = 297$. {\sl Right}: growing mode spectrum for ${\rm Rm} = 10^5$ (multiplied by 30 for convenience).
    }
 \end{description}
\end{figure}

Half-widths of the spectra of Fig.\,6 coincide in the order of magnitude with positions of their maxima. With the correlation function normalised as $B(0) =1$, the spectrum amplitude is proportional to ${\rm Re}^{3/4}$.

\begin{figure}[htb]
 \begin{center}
 \includegraphics[width=12.0 cm]{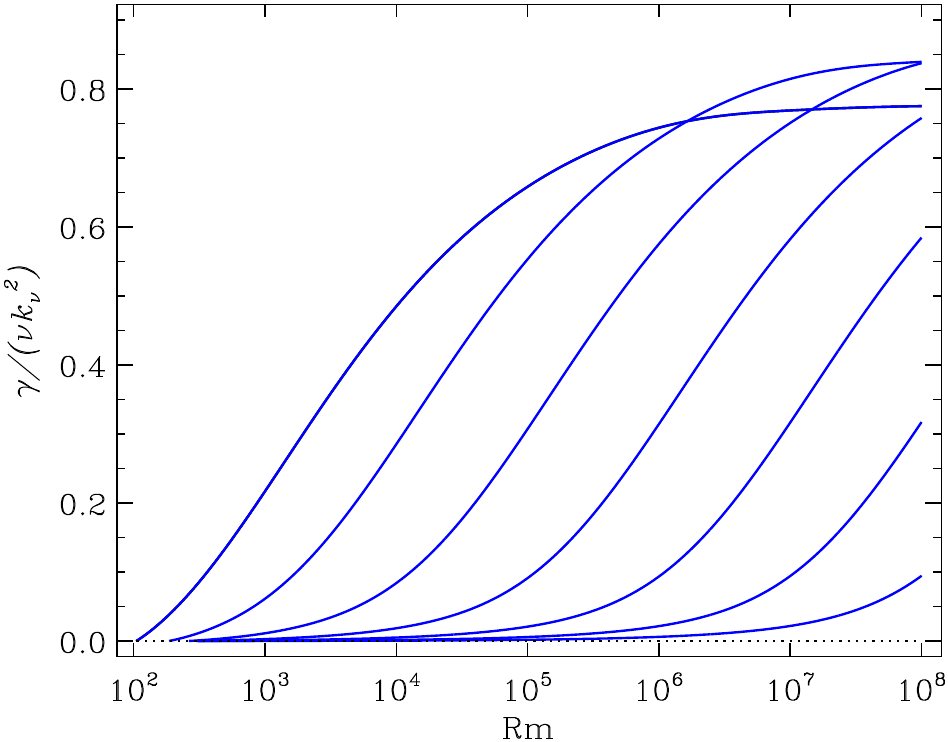}
 \end{center}
 \begin{description}
 \item{\small {\bf Fig.\,8.} Normalized growth rate $\gamma/(\nu k_\nu^2)$ as functions of $\rm Rm$ for a fixed Reynolds number. Reynolds number is constant for each line in the plot but varies as $10^N$ with $N = 2,3,...,8$ from left to right line.
    }
 \end{description}
\end{figure}

As $\rm Rm$ increases at a fixed Reynolds number, the spectrum of the most rapidly growing mode shifts in the direction of increasing wave number $k$. This is clearly seen in Fig.\,7, where the spectra for the threshold value $\rm Rm_c$ and $\rm Rm = Re$ are compared (note that the spectra (\ref{20}) for Kazantsev equation solutions are positive-definite, as they should be). As $\rm Rm$ grows beyond the Reynolds number, the spectral displacement to larger $k$ stops. This is probably because of the lack of sufficiently intensive motions for magnetic field generation at $k > k_\nu$. 

This stop-of-shift is consequential to the growth rate dependence on the problem parameters. The growth rates cannot be larger than the inverse lifetime $1/\tau_k$ of the field generating eddies. The smallest lifetimes, $\tau_{k_\nu} \simeq 1/(\nu k_\nu^2)$, have the eddies of dissipative scales. Accordingly, the normalised growth rates of Fig.\,8 are smaller than one.

Magnetic field generation at small magnetic Prandtl numbers settles at the Ohmic dissipation scale (Figs~6 and 7). Two competing processes - field amplification and diffusive decay - control growth rates in this regime. Rates of both processes ($\sim 1/\tau_k$) coincide in their order of magnitude. The growth rates for small $\rm Pm$ are therefore small $\gamma \propto \ln ({\rm Rm}/{\rm Rm_c})$ [\ref{KR12}]. As $\rm Pm$ approaches the value of one, diffusion weakens and the growth rate increases faster with $\rm Rm$. However, unlimited growth of $\gamma \propto {\rm Rm}^{1/2}$ discussed in the literature is not found. A further increase in ${\rm Pm} > 1$ leads to a saturation of $\gamma$ at a constant value somewhat below $1/\tau_{k_\nu}$.
 \bl
 \noindent{\bf 5. Conclusions and perspectives}

\noindent
Our results can be explained and interpreted as follows. 

The small-scale dynamo is controlled by two competing processes: field amplification due to field-line stretching and diffusion (Ohmic and turbulent). The turbulent diffusion transmits magnetic energy to increasingly small scales where the Ohmic dissipation eventually comes into effect. The rate of field amplification $Q(r)$ (Fig.\,3) and the rate of turbulent fragmentation of scales $\eta_{_{\rm T}}(r)/r^2$ (Fig.\,2) both increase with decreasing scale $r$ until $r$ falls to the viscous scale $1/k_\nu$. Ohmic dissipation rate $\eta/r^2$ grows with decreasing $r$ faster than the rate of turbulent diffusion and exceeds the latter for $r < 1/k_\eta$ where Ohmic diffusion stops the field amplification. The largest rate of magnetic field generation is achieved in the spectrum region of $k \sim k_\eta$ (Figs~6 and 7). 

As the magnetic Prandtl number increases to ${\rm Pm} \sim 1$ and further, the field generation range shifts to $k \sim k_\nu$, where the turbulent fragmentation of scales weakens because of the absence of eddies of smaller scale in the turbulence spectrum and the magnetic energy spectrum does not shift to smaller $k$ any further. Meanwhile, the growth rates of Fig.\,8 increase strongly. Due to a decline in Ohmic diffusion, the increase in $\gamma$ proceeds for ${\rm Pm} > 1$, but it is limited by the value $\gamma_{\rm max} \simeq 1/\tau_{k_\nu}$ of the inverse lifetime of the smallest eddies. Our numerical method is confined to the Reynolds numbers ${\rm Re}, {\rm Rm} \leq 10^8$. Under this restriction, the growth rate saturation in Fig.\,8 is evident for ${\rm Re} < 10^4$ only. The value of $\gamma_{\rm max}$ increases slightly with the Reynolds number. This is probably because of a similar increase in the field amplification coefficient of Fig.\,3. 

The case of ${\rm Pm} \ll 1$ refers to the sun and other stars with convective turbulence. The Kazantsev model shows the small-scale dynamos for this case. But the cor\-res\-pon\-ding growth rates are small. This causes a problem: the dynamo effect results from a slight predominance of field amplification over diffusive decay in a model whose accuracy is uncertain. Verification of the model with observational or experimental data seems to be pertinent. 

The Kazantsev model is considered to be kinematic, i.e., relevant to the case of  magnetic energy being small compared to the turbulent kinetic energy. In the nonlinear stage of the dynamo, the field amplification is stabilised by a reaction of the magnetic field back on the flow [\ref{YLS19}]. Application of the Kazantsev equation to this stage is, however, limited by uncertainty in the results of this back reaction only. The limitation can be repealed if correlation functions of both magnetic field and turbulent flow are simultaneously known for some range of scale $r$. A substitution of the correlation functions in the right side of Eq.\,(\ref{2}) would then evaluate the accuracy of the equation (the left side has to be zero in the nonlinear regime). Note that the correlation functions for magnetic fields and flows are simultaneously accessible to solar observations [\ref{AYW01},\ref{A17}].

A perspective of the numerical method of this paper could also be an allowance for turbulent hydrodynamic helicity in the small-scale dynamo problem. 
\bl
{\bf Funding.} The work is financially supported by the Ministry of Science and High Education of the Russian Federation.
\bl
{\bf Conflict of interest.} The author declares no conflict of interest.
\bll
{\bf References}
\begin{enumerate}
\item\label{ZRS83} Ya.\,B.~Zeldovich, A.\,A.~Ruzmaikin, and
    D.\,D.~Sokoloff, {\sl Magnetic Fields in Astrophysics}, Gordon and
    Breach Science Publ., New York (1983).
\item\label{KR84} F.~Krause and K.-H.~R\"adler, {\sl Mean-Field Magnetohydrodynamics and Dynamo Theory}, Akademie-Verlag, Berlin (1980).
\item\label{RKH13} G.~R\"udiger, L.\,L.~Kitchatinov, and
    R.~Hollerbach, {\sl Magnetic Processes in Astrophysics: Theory,
    Simulations, Experiments}, Wiley-VCH Verlag GmbH, Weinheim
    (2013).
\item\label{K67} A.\,P.~Kazantsev, Sov. Phys. JETP {\bf 26}, 1031 (1968).
\item\label{NRS83} V.\,G.~Novikov, A.\,A.~Ruzmaikin, and D.\,D.~Sokolov, Sov.
    Phys. JETP {\bf 58}, 527 (1983).
\item\label{Cea99} M.\,Chertkov, G.\,Falkovich, I.\,Kolokolov, and
    M.\,Vergassola, Phys. Rev. Lett. {\bf 83} 4065 (1999).
\item\label{BC04} S.~Boldyrev and F.~Cattaneo, Phys. Rev. Lett. {\bf
    92} 144501 (2004).
\item\label{BCR05} S.~Boldyrev, F.~Cattaneo, and R.~Rosner, Phys.
    Rev. Lett. {\bf 95}, 255001 (2005).
\item\label{BSS13} S.~Bovino, D.\,R.\,G.~Schleicher, and J.~Schober,
    New J. Phys. {\bf 15} 013055 (2013).
\item\label{Nea22} N.~Kriel, J.\,R.~Beattie, A.~Seta, and
    C.~Federrath, MNRAS {\bf 513} 2457 (2022).
\item\label{MYS25} I.\,V.~Makarova, Е.\,V.~Yushkov, D.\,D.~Sokoloff, Zh.\,Exp.\,Teor.\,Fiz. {\bf 168}, 843 (2025).
\item\label{Kea22} A.\,V.~Kopyev, A.\,M.~Kiselev, A.\,S.~Il'yn,
    V.\,A.~Sirota, and K.\,P.~Zybin, Astrophys. J. {\bf 927}, 172
    (2022).
\item\label{Kea24} A.\,V.~Kopyev, A.\,S.~Il'yn, V.\,A.~Sirota, and
    K.\,P.~Zybin, MNRAS {\bf 527}, 1055 (2024).
\item\label{MB10} L.\,M.~Malyshkin and S.\,Boldyrev, Phys. Rev. Lett.
    {\bf 105}, 215002 (2010).
\item\label{KR12} N.~Kleeorin, I.\,Rogachevskii, Phys. Scripta
    {\bf 86}, 018404 (2012).
\item\label{MY67} A.\,S.~Monin and A.\,M.~Yaglom,
    {\sl Statistical Fluid Mechanics}, ed. J.~Lumley, MIT Press, Cambridge
    (1975).
\item\label{VK86} S.\,I.~Vainshtein and L.\,L.~Kichatinov, J. Fluid
    Mech. {\bf 168}, 73 (1986).
\item\label{K89} L.\,L.~Kichatinov, J. Fluid Mech. {\bf 208}, 115
    (1989).
\item\label{K41} A.~Kolmogorov, DoSSR {\bf 30}, 301 (1941).
\item\label{Mea97} D.\,O.~Martinez, S.~Chen, G.\,D.~Doolen,
    R.\,H.~Kraichnan, L.-P.~Wang, and Y.~Zhou, J. Plasma Phys. {\bf
    57}, 195 (1997).
\item\label{R26} L.\,F.~Richardson, Proc. Roy. Soc. London. Ser.\,A
    {\bf 110}, 709 (1926).
\item\label{O41} A.~Obukhov, Izv. Akad. Nauk SSSR. Ser.
    Geogr. Geofiz. {\bf 5}, 453 (1941).
\item\label{Pea92} W.\,H.~Press, S.\,A.~Teukolsky,
    W.\,T.~Vetterling, and B.\,P.~Flannery, {\sl Numerical Recipes. The
    Art of Scientific Computing}, Cambridge Univ. Press (1992).
\item\label{Sea04} A.\,A.~Schekochihin, S.\,C.~Cowley, J.\,L.~Maron,
    and J.\,C.~McWilliams, Phys. Rev. Lett. {\bf 92}, 054502 (2004).
\item\label{Iea07} A.\,B.~Iskakov, A.\,A.~Schekochihin,
    S.\,C.~Cowley, J.\,C.~McWilliams, M.\,R.\,E.~Proctor, Phys. Rev.
    Lett. {\bf 98}, 208501 (2007).
\item\label{YLS19} E.\,V.~Yushkov, A.\,S.~Lukin, and
    D.\,D.~Sokoloff, JETP {\bf 129}, 1086 (2019).
\item\label{AYW01} V.~Abramenko, V.~Yurchishin, and H.~Wang, Sol.
    Phys. {\bf 201}, 225 (2001).
\item\label{A17} V.\,I.~Abramenko, MNRAS {\bf 471}, 3871 (2017) 
\end{enumerate}

\end{document}